\documentclass[aps,prd,preprintnumbers,
showpacs,showkeys,nofootinbib,
superscriptaddress,fleqn,floatfix,
tightenlines,twocolumn,10pt]{revtex4-2}
\usepackage{amsmath,amsfonts,amssymb,amscd,amsxtra,amsthm}
\usepackage{graphicx}  
\usepackage{dcolumn}  
\usepackage{bm}         
\usepackage{slashed}
\usepackage{cancel}
\usepackage{color}
\usepackage{float}
\usepackage{mathtools}
\usepackage{amsbsy} 
\usepackage{amstext}
\usepackage{multirow}
\newcolumntype{L}{>{\raggedright\arraybackslash}X}
\usepackage{ltablex}
\usepackage{siunitx}

\usepackage[utf8]{inputenc}
\usepackage[normalem]{ulem} 
\usepackage[dvipsnames]{xcolor} 
\renewcommand\sout{\bgroup \color{red} \ULdepth=-.5ex \ULset}

\makeatletter

\begin{document}
\preprint{INHA-NTG-07/2022}
\title{Instanton effects on electromagnetic transitions of charmonia}  
\author{Ki-Hoon~Hong}
\email{kihoon@inha.edu}
\affiliation{Department of Physics, Inha University, Incheon 22212,
 Republic of Korea}
 \author{Hyun-Chul~Kim}
 \email{hchkim@inha.ac.kr}
 \affiliation{Department of Physics, Inha University, Incheon 22212,
  Republic of Korea}
 \affiliation{School of Physics, Korea Institute for Advanced Study
   (KIAS), Seoul 02455, Republic of Korea}
\author{Ulugbek Yakhshiev}
\email[E-mail: ]{yakhshiev@inha.ac.kr}
\affiliation{Department of Physics, Inha University, Incheon 22212,
 Republic of Korea}

\begin{abstract}
We investigate the mass spectrum and electromagnetic transitions 
of charmonia, emphasizing the instanton effects on them. 
The heavy-quark potential consists of the Coulomb-like potential
from one-gluon exchange and the linear confining potential.    
We introduce the nonperturbative heavy-quark potential derived from
the instanton vacuum. We also consider the screened confining
potential, which better describes the electromagnetic
decays of higher excited states. Using this improved heavy-quark
potential, we compute the mass spectrum and electromagnetic decays of
the charmonia. Focusing on the instanton effects, we discuss the
results compared with the experimental data and those from other
works. The instanton effects are marginal on the electromagnetic
decays of charmonia. 
\end{abstract}


\date{\today}
\maketitle
\section{Introduction}
The quarkonium is the simplest and yet crucial system that consists of 
one heavy quark and one heavy antiquark among all hadrons. 
Since it is very heavy, a nonrelativistic (NR) approach is suitable for
describing its structure and properties. Moreover, the accurate
measurements of masses and radiative decay widths for the low-lying
quarkonia provide a precision test of any theory based on perturbative
and non-perturbative quantum chromodynamics (see reviews on
quarkonia~\cite{Eichten:2007qx, Voloshin:2007dx, Brambilla:2010cs,
  Patrignani:2012an}). The quarkonia have successfully been described  
by quantum mechanical potential models~\cite{Eichten:1974af,
  Eichten:1978tg}. The static heavy-quark potential consists of two
main contributions: The Coulomb-like term and the phenomenological
quark-confining one. The former arises from one-gluon
exchange (OGE) between a heavy quark ($Q$) and a heavy antiquark 
($\overline{Q}$), based on perturbative quantum chromodynamics
(pQCD)~\cite{Susskind:1976pi, Appelquist:1977tw, 
  Appelquist:1977es, Fischler:1977yf}. Higher-order corrections from
pQCD were also considered~\cite{Peter:1996ig, Peter:1997me,
  Schroder:1998vy, Smirnov:2009fh, Anzai:2009tm}. Since OGE comes from
pQCD, the Coulomb-like potential governs the short-range dynamics
inside charmonia. It should fade away as the distance between heavy
and anti-heavy quarks and then will be taken over by the
effects of the quark confinement~\cite{Wilson:1974sk}. The heavy-quark
potential for the quark confinement can be derived phenomenologically
from the Wilson loop, which increases monotonically as the distance
between heavy quarks increases~\cite{Eichten:1974af, Eichten:1978tg}.   

While the Coulomb-like and quark-confining terms constitute the
main contributions to the static heavy-quark potential, there is still 
a nonperturbative effect that arises from the instanton vacuum. 
The nonperturbative heavy-quark potential was derived from the
instanton vacuum~\cite{Diakonov:1989un} and was examined by computing 
the charmonium spectra~\cite{Yakhshiev:2016keg,
  Yakhshiev:2018juj}. While the instanton effects on the heavy-quark
potential are small, they provide significant physical
implications. Firstly, the heavy quark acquires an additional
dynamical mass from the instantons, i.e. $\Delta M_I\simeq 70$
MeV~\cite{Diakonov:1989un, Yakhshiev:2016keg,
  Yakhshiev:2018juj,Musakhanov:2021gof}, which allows one to use the
value of the heavy-quark mass close to the physical one in the
heavy-quark potential. Secondly, an additional contribution from the
instantons makes it possible to employ the value of the strong
coupling constant near the charm-quark mass scale. Note 
that often the strong coupling constant used in the Coulomb-like
potential was overestimated~\cite{Godfrey:1985xj, Barnes:2005pb,
  Deng:2016stx}. In addition to the instanton effects, we want to
modify the linear confining potential. When the $1/m_Q$ corrections
are considered where $m_Q$ is the heavy-quark mass, the heavy quark is
no more static. The light quark-antiquark pair will create from the 
vacuum at a certain scale ($\sim 1$ fm). Furthermore, the created
quark-antiquark pair will screen the color
charge~\cite{Laermann:1986pu, Born:1989iv}. The screened confining 
potential has been adopted to describe the electromagnetic (EM) decays
of charmonia~\cite{Li:2009zu, Deng:2016stx}. While the EM transitions
of the charmonia have extensively been studied within various theoretical
frameworks such as the heavy-quark potential
models~\cite{Barnes:2003vb, Barnes:2005pb, Deng:2016stx}, lattice
QCD~\cite{Li:2009zu, Dudek:2006ej, Dudek:2009kk, Chen:2011kpa,
  Yang:2012mya, Donald:2012ga, Becirevic:2014rda}, QCD sum
rules~\cite{Beilin:1984pf, Guo:2019xqa, Li:2020rcg}, Bethe-Salpeter 
equations~\cite{Bhatnagar:2020vpd, Guleria:2021fpw}, 
potential NR QCD (pNRQCD)~\cite{Brambilla:2005zw,
  Pineda:2013lta}, and quark models~\cite{Ke:2013zs, Guo:2014zva,
  Shi:2016cef, Li:2018uif, Ganbold:2021nvj}, the EM decays of higher
excited charmonium states are not fully understood theoretically.  

In the present work, we want to extend the previous works and examine
the instanton effects on the charmonium spectrum and EM transitions of
the charmonia. We sketch the current work as follows: In 
Section II, we briefly explain how the heavy-quark potential is
derived from the instanton vacuum. In Section III, we reproduce the
mass spectrum of the charmonia with a new type of the confining
potential introduced. In Section IV, we derive the spin-dependent part
of the heavy-quark potential. In Section V, we 
discuss the results for the charmonium spectrum and EM decay
rates. The final Section is devoted to the summary and conclusion. 

\section{Heavy-quark potential from the instanton
  vacuum}\label{sec:HQP} 
We begin by recapitulating the derivation of the nonperturbative
heavy-quark potential from the instanton vacuum. 
\subsection{Non-perturbative heavy-quark  potential }
\begin{figure}[H]\centering
\includegraphics[width=7cm]{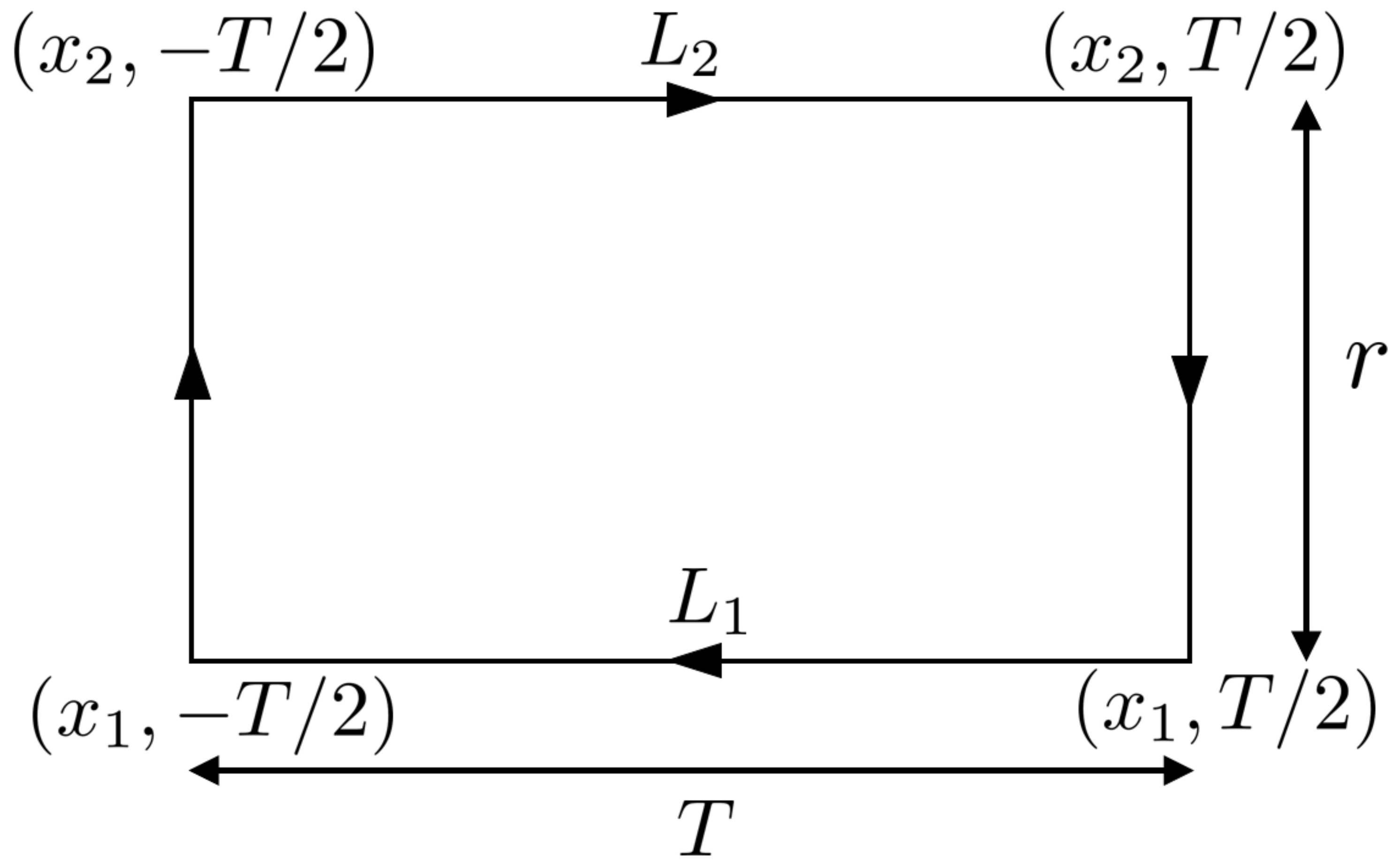}
\caption{The rectangular Wilson loop along a contour $T\times r$ with
  $T\rightarrow\infty$. In this limit, one can ignore the spatial line
  because $r$ is very small compared to $T$.} 
\label{fig:1}
\end{figure}

The heavy-quark potential can be defined by the Wilson loop as shown
in Fig.~\ref{fig:1}.  
Diakonov et al.~\cite{Diakonov:1989un} derived the central part of the
nonperturbative heavy-quark potential, regarding an instanton packing
parameter $\lambda=\bar{\rho}^4/R^4(\sim 0.01)$ as an expansion
parameter, where $\bar{\rho}$ denotes the average instanton size and
$R$ the instanton inter-distance that is represented as $R=N/V$. In
the instanton liquid model, the number of instanton $N$ and the four
dimensional volume $V$ give $N/V\simeq(200\,\rm MeV)^4$ and
$\bar{\rho}\simeq(600\,\mathrm{MeV})^{-1} \simeq
1/3\,\mathrm{fm}$~\cite{Shuryak:1982hk, Diakonov:1983hh,
  Diakonov:1985eg}. The gluon field $\mathcal{A}_\mu(x)$ in Euclidean
space can be decomposed into the classical($A$) and quantum($B$) parts  
\begin{align}
    \mathcal{A}_\mu(x)=A_\mu(x,\xi)+B_\mu(x),
\end{align}
where the classical background field $A_\mu(x,\xi)$ is given by the
summation over all (anti)instanton fields  
$A_\mu(x,\xi)=\sum_iA_{i\mu}(x,\xi_i)$. Here, $\xi_i=(z_i,U_i,\rho_i)$
denote collective coordinates: the instanton positions $z_i$,
the color orientations $U_i$, and the sizes of the instantons 
$\rho_i$. In the large $N_c$ (number of color) limit, the width of the
instanton distribution is of order $1/N_c$, so the instanton
distributions can be approximated by the delta functions. Thus,
$\rho_i$ is just the average size of the instanton $\bar{\rho}$,
$\rho_i=\bar{\rho}$. The instanton field $A_{\pm,\mu}$ in the singular
gauge is expressed as 
\begin{align}
    A_{\pm,\mu}(x,z_\pm) = \frac{\eta^{\mp}_{\mu\nu}(x-z_\pm)_\mu
  \lambda^a \bar{\rho}^2}{(x-z_\pm)^2((x-z_\pm)^2+\bar{\rho}^2)},  
\end{align}
where $\eta_{\mu\nu}^\mp$ stand for the 't Hooft symbols. 
\begin{align}
  \eta_{a\mu\nu}^{\pm}=\left\{\begin{array}{cc}
    \epsilon_{a\mu\nu}\quad&:\ \mu,\nu=1,2,3\\
    \mp\delta_{a\nu}\quad&:\ \mu=4\\
    \pm\delta_{a\mu}\quad&:\ \nu=4\\
    0\quad&:\ \mu,\nu=4
  \end{array}\right.
\end{align}
Here, the $+$ and $-$ signs designate respectively the instanton and
the antiinstanton. Note that the heavy-quark and antiquark correlator
is just the same as the Wilson
loop~\cite{Musakhanov:2020hvk}. Considering that the instanton medium
is dilute, we can average over the pesudoparticles (instantons and
antiinstantons) independently. Thus, the average of the Wilson loop
over the pseudoparticles can be expressed as  
\begin{align}
  w\equiv\langle\kern-2\nulldelimiterspace \langle W
  \rangle\kern-2\nulldelimiterspace \rangle &=\int
 d \xi\frac{1}{D^{(1)}}     \frac{1}{D^{(2)}},
  \label{eq:QQbarCol}
\end{align}
where $d\xi =V^{-1}\prod_{i=1}dz_idU_i$ stands for the measure of
the zero modes and $D^{(i)}$ is the inverse of the heavy-quark
propagator expressed by $D^{(i)}=\theta^{-1}-\sum_Ia_I^{(i)}$. The
double angle brackets represent the ensemble average over the  
pesudoparticles~\cite{Diakonov:1989un}. $\theta$ designates the
inverse of the time derivative $d/dt$: $\langle
t|\theta|t'\rangle=\theta(t-t')$ with the Heaviside step function
$\theta(t-t')$. $a_I^{(i)}$ represent
$a_I^{(i)}=iA_{I,\mu}^{(i)}(\xi)\dot{x}_\mu(t)$, which is aligned
along the tangential direction to the Wilson line. The superscripts  
in Eq.~\eqref{eq:QQbarCol} denote the corresponding Wilson lines as 
shown in Fig.~\ref{fig:1}. After the integral over the
color-orientations and using the Pobylitsa
equation~\cite{Pobylitsa:1989uq}, we obtain the inverse of the Wilson 
loop in powers of $N/VN_c$ 
\begin{align}
&w^{-1}=\theta^{-1}\theta^{-1} + \frac{N}{2VN_c} \mathrm{Tr}_c
\sum_{\pm}\int d^4z_{\pm}\cr    
& \times\left[\theta^{-1} \theta^{-1}- \theta^{-1}w_\pm^{(1)}
  \theta^{-1} \theta^{-1}w_{\pm}^{(2)}\theta^{-1}\right]\cr
& +\mathcal{O}\left(\left(\frac{N}{2VN_c}\right)^2\right),
    \label{eq:NPV}
\end{align}
where $\mathrm{Tr}_c$ is the trace over color space. $w_{\pm}$ is
defined as 
\begin{align}
w_{\pm}=(\theta^{-1}-a_{\pm})^{-1}.
\end{align}

Performing the Fourier transformation~\cite{Diakonov:1989un}, we
obtain 
\begin{align}
    \langle
  t_1^{(1)}|w|t_1^{(2)}\rangle =
  \int\frac{d\omega}{2\pi}e^{i\omega(t_1^{(1)}-t_1^{(2)})}
  \frac{1}{w^{-1}(\omega)},   
    \label{eq:FTW}
\end{align}
where $t_1^{(1)}$ and $t_1^{(2)}$ are $-T/2$ in $L_1$ and $-T/2$ in
$L_2$, respectively. In the infinite time limit $T\rightarrow \infty$,
the heavy quark potential is defined by the Wilson loop as follows 
\begin{align}
    \langle t_1^{(1)}|w|t_1^{(2)}\rangle
    &\approx\exp\left[-V_I T\right],
\end{align}
where $V_I$ denotes the instanton-induced potential,
respectively. Then $V_I$ is derived as  
\begin{align}
    V_I(r) &=\frac{N}{2VN_c}\sum_\pm \int d^3z_\pm\cr 
    &\times       \mathrm{Tr}_c\left[1-P\exp 
\left(i\int_{-\infty}^{\infty}dx_4A_{\pm,4}\right)\right.\cr 
    &\qquad\qquad\times \left.P\exp
\left(i\int_{-\infty}^\infty dy_4A_{\pm,4}\right)\right]
\label{eq:VNP}
\end{align}

\subsection{Spin-dependent potential}
So far, we have obtained the static heavy-quark potential from the
instanton vacuum. However, we need to consider the spin-dependent
correction to obtain the mass splitting in charmonia. In this section
we will briefly show how the spin-dependent parts can be
constructed from the centeral Coulomb-like and confining
potentials. We can obtain them by using the Fermi-Breit
equation~\cite{Eichten:1980mw, Voloshin:2007dx}   
\begin{align}
    V_{SD}(r)=&V_{SS}(r)\bm{S}_Q\cdot\bm{S}_{\bar{Q}} +
                V_{LS}(r)\bm{L}\cdot\bm{S}\cr  
    +&V_T(r)[3(\bm{S}_Q\cdot\hat{n})(\bm{S}_{\bar{Q}}
       \cdot\hat{n})-\bm{S}_Q\cdot\bm{S}_{\bar{Q}}], 
\label{eq:sdpot}
\end{align}
where the spin-dependent potential $V_{SD}$ can be expressed by three 
terms, i.e., the spin-spin interaction $V_{SS}$, the spin-orbit
interatcion $V_{LS}$ and the tensor interaction $V_T$. The radial
parts of Eq.~\eqref{eq:sdpot} are expressed as 
\begin{align}
    V_{SS}(r)&=\frac{2}{3m_Q^2}\nabla^2 V_{\rm
               v}=\frac{32\pi\alpha_s}{9m_Q^2}
               \delta(r),\label{eq:VSS}\\  
    V_{LS}(r)&=\frac{1}{2m_Q^2r}\left(3\frac{dV_{\rm v}}{dr} -
               \frac{dV_{\rm s}}{dr}\right),\label{eq:VLS}\\ 
    V_T(r)&=\frac{1}{3m_Q^2}\left(\frac{1}{r}\frac{dV_{\rm v}}{dr}
            -\frac{d^2V_{\rm s}}{dr^2}\right),\label{eq:VT} 
\end{align} 
where $V_{\rm v}$ and $V_{\rm s}$ denote the Coulomb-like potential
and the confining potential, respectively. The spin-dependent
potentials appear from the next-to-leading order in powers of
$1/m_Q^2$. The Dirac delta function $\delta$ in Eq.~\eqref{eq:VSS}
should be smeared by the Gaussian form to avoid a singular behavior of 
the spin-spin potential  
\begin{align}
  \tilde{\delta}(r)=\left(\frac{\sigma}{\sqrt{\pi}}\right)^3e^{-\sigma^2r^2},  
\end{align}
where $\sigma$ is a smearing parameter.
 
The instanton-induced spin-dependent potential is derived from the
Wilson line in the next-to-leading order of $1/m_Q^2$. The Wilson line
in Fig.~\ref{fig:1} is expressed as 
\begin{align}
    W(x&,y;A)\cr
    &=W_0(x,y;A)-\int d^4zW_0(x,z;A)\cr
    &\quad\times\left[i\slashed{D}_\perp \frac{1}{iv\cdot
      D+2m_Q}i\slashed{D}_\perp\right]W(z,y;A),\label{eq:NLW} 
\end{align}
where
\begin{align}
    W_0(x,y;A)=P\exp\left(i\int_{x_4}^{y_4}A_4dz_4\right)
  \delta^{(3)}(\bm{x}-\bm{y}). 
\end{align}
In the rest frame, the four-velocity $v_\mu$ is written as 
\begin{align}
    v_\mu=(1,\boldsymbol{0}),\quad \slashed{D}_\perp =
  \gamma^iD_i,\quad (i=1,2,3), 
\end{align}
where $D_\mu$ is a covariant derivative that is defined as 
$D_\mu=\partial_\mu-igA_\mu$. If we expand the inside of the square
bracket of Eq.~\eqref{eq:NLW} in powers of $1/m_Q$ then  
\begin{align}
    &i\slashed{D}_\perp\frac{1}{iv\cdot D+2m_Q}i\slashed{D}_\perp\cr
        &=i\slashed{D}_\perp\left(\frac{1}{2m_Q}-\frac{i}{4m_Q^2}v\cdot
          D+\mathcal{O}\left(\frac{1}{m_Q^3}\right)\right)i\slashed{D}_\perp. 
\label{eq:18}  
\end{align}
In Euclidean space, one can obtain the following expressions 
\begin{align}
    (i\slashed{D}_\perp)^2&=-\bm{D}^2+\bm{\sigma}\cdot\bm{B},\cr
    i\slashed{D}_\perp(iv\cdot D)i\slashed{D}_\perp&=i\bm{E}
    \cdot\bm{D}+\bm{\sigma}\cdot(\bm{E}\times\bm{D}),
\label{eq:SBP} 
\end{align}
where $\boldsymbol{D}$ and $\boldsymbol{B}$ are called chromoelectric
and chromomagnetic fields, respectively. The component of the gluon
field strength tensor ($G_{0i}$) and the spatial component of the
covariant derivative ($D_i$) in Euclidean space are given as  
\begin{align}
    G_{0i}&=E_i=iG_{{\rm E},4i}=-iE_{{\rm E},I},\cr
    D_i&=-D_{{\rm E},i}.
\end{align}
We omit the Euclidean symbol E from now on. Hence the Wilson line can
be iteratively expressed in powers of $1/m_Q$: 
\begin{align}
    &W(x,y;A)\cr
    &=W_0(x,y;A)\cr
    &-\frac{1}{2m_Q}\int d^4zW_0(x,z;A)(-\bm{D}^2+\boldsymbol{\sigma}
      \cdot\bm{B})W_0(z,y;A)\cr 
    &-\frac{1}{4m_Q^2}\int d^4z W_0(x,z;A)\cr
    &\qquad\times(-i\bm{E}\cdot\bm{D}-\boldsymbol{\sigma}
      \cdot(\bm{E}\times\bm{D}))W_0(z,y;A)\cr 
    &+\frac{1}{4m_Q^2}\int d^4zd^4z'W_0(x,z;A)\cr
    &\qquad\times(-\bm{D}^2+\boldsymbol{\sigma}
      \cdot\bm{B})W_0(z,z';A)\cr 
    &\qquad\times(-\bm{D}^2+\boldsymbol{\sigma}
      \cdot\bm{B})W_0(z',y;A)\theta(z_4'-z_4)\cr 
    &+\mathcal{O}\left(\frac{1}{m_Q^3}\right).
\end{align}
Having carried out lengthy calculations (see the
details in Ref.~\cite{Eichten:1980mw,Yakhshiev:2016keg}), we derive
the spin-dependent potential as follows: 
\begin{align}
    V_{SD}^I(r)&=\frac{1}{4m_Q^2}(\boldsymbol{\sigma}_1
                 \cdot\bm{L}_1-\boldsymbol{\sigma}_2\cdot\bm{L}_2)\cr 
    &\quad\times\left(\frac{1}{r}\frac{\partial V_I(r)}{\partial r}
      +\frac{2}{r}\frac{\partial V_1(r)}{\partial r}\right)\cr 
    &+\frac{1}{2m_Q^2}(\boldsymbol{\sigma}_2\cdot\bm{L}
      -\boldsymbol{\sigma_1}\cdot\bm{L}_2)\frac{1}{r}\frac{\partial
      V_2(r)}{\partial r}\cr 
    &+\frac{1}{12m_Q^2}\{(3\boldsymbol{\sigma}_1\cdot\hat{r})
      (\boldsymbol{\sigma}\cdot\hat{r})
      -\boldsymbol{\sigma}_1\cdot\boldsymbol{\sigma}_2\}V_3(r)\cr  
    &+\frac{1}{12m_Q^2}\boldsymbol{\sigma}_1\cdot\boldsymbol{\sigma}_2V_4(r),
\end{align}
where
\begin{align}
    V_1(r)&=-V_2(r)=-\frac{1}{2}V_I(r),\cr
    V_3(r)&=\frac{\partial^2V_I(r)}{\partial r^2} -
            \frac{1}{r}\frac{\partial V_I(r)}{\partial r},\cr 
    V_4(r)&=\boldsymbol{\nabla}^2V_I(r).
\end{align}
Then we can rewrite the spin-dependent part of the instanton-induced 
potential as 
\begin{align}
    V_{SD}^I(r)&=V_{SS}^I(r)\bm{S}_Q\cdot
                 \bm{S}_{\bar{Q}}+V_{LS}^I(r)\bm{L}\cdot\bm{S},\cr 
    &+V_T^I(r)[3(\bm{S}_Q\cdot\hat{n})(\bm{S}_{\bar{Q}}
      \cdot\hat{n})-\bm{S}_Q\cdot\bm{S}_{\bar{Q}}], 
\end{align}
where the radial parts are expressed as 
\begin{align}
    &V_{SS}^I(r)=\frac{1}{3m_Q^2}\nabla^2V_I(r)
    \label{eq:VSSI}\\
    &V_{LS}^I(r)=\frac{1}{2m_Q^2}\frac{1}{r}\frac{dV_I(r)}{dr}\\
    &V_{T}^I(r)=\frac{1}{3m_Q^2}\left(\frac{d^2}{dr^2}
      -\frac{1}{r}\frac{d}{dr}\right)V_I(r). 
\end{align}
\section{Mass spectrum of the charmonia}
\begin{figure}[h]
  \centering
  \includegraphics[width=8cm]{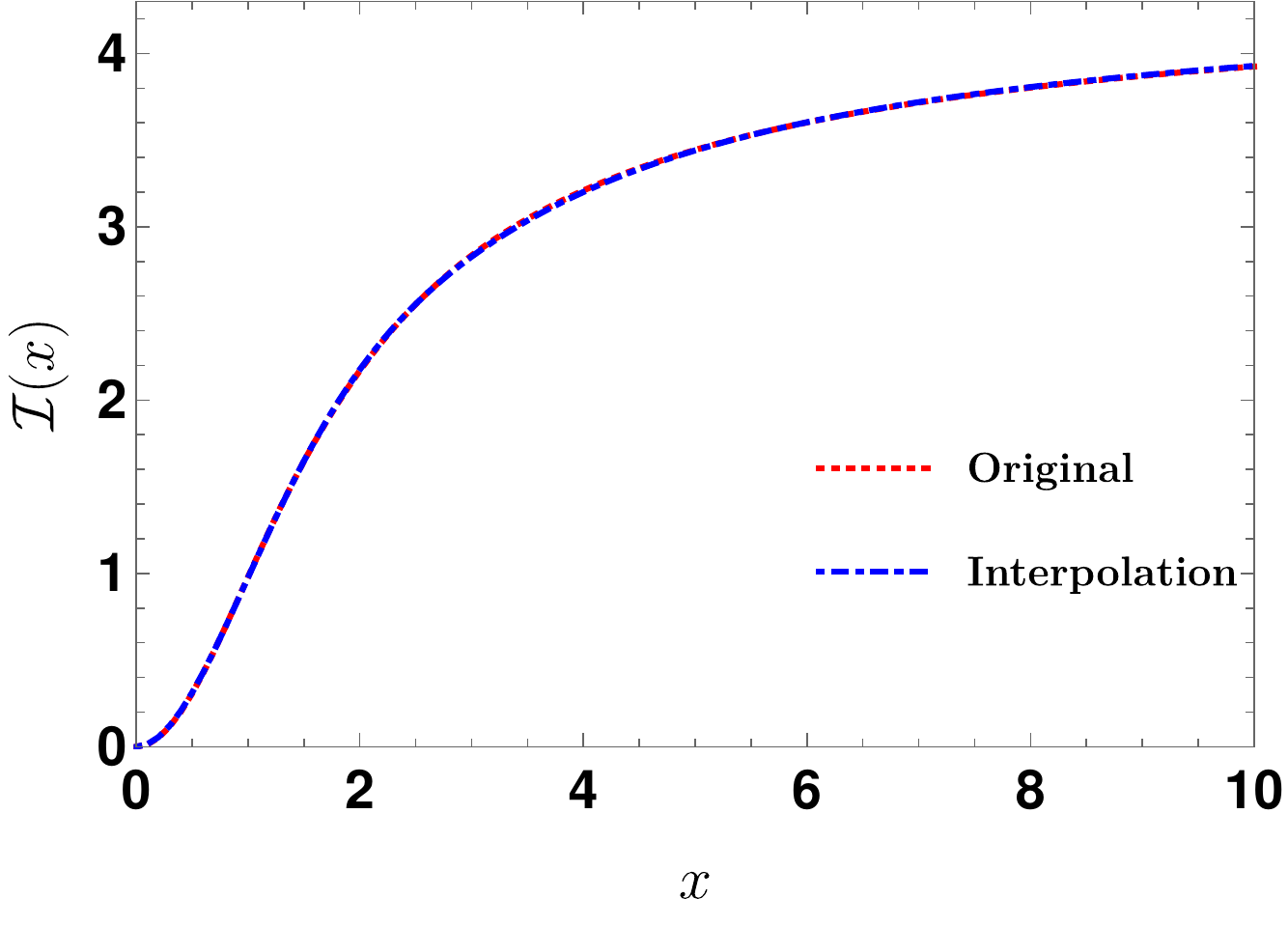}
  \caption{The dotted curve depicts the numerical result for the
    original integral function whereas the dot-dashed line draws
    the interpolating function $\mathcal{I}(x)$. 
    Both $x$ and 
    $\mathcal{I}(x)$ are dimensionless. $\mathcal{I}(\infty)$
    converges to 4.41625.} 
  \label{fig:2}
\end{figure}

Recently, the mass spectrum of the charmonia is studied within the
present framework~\cite{Yakhshiev:2018juj, Pandya:2018jxw,
  DSouza:2018iyc, Dorokhov:2005cw}. Especially, 
in Ref.~\cite{Yakhshiev:2018juj}, 
Although the instanton effects are small on the mass spectrum of the
quarkonium, they allow one to use the value of $\alpha_s$ closer to
the physical one than those in other potential approaches, as
mentioned previously. One can also use the smaller value of the
heavy-quark mass to describe the mass spectrum. 
The instanton-indueced potential given in Eq.~\eqref{eq:VNP} is
explicitly written as the following integral  
\begin{align}
    V_{I}(r)&=\frac{4\pi\lambda}{N_c\bar{\rho}}\int_0^\infty
              y^2dy\int_{-1}^1 dt\cr 
    &\times\left[1-\cos\left(\frac{\pi y}
      {\sqrt{y^2+1}}\right)\right.\cr 
    &\quad\times\cos\left(\pi\sqrt{
      \frac{y^2+x^2+2xyt}{y^2+x^2+2xyt+1}}\right)\cr 
    &-\frac{y+xt}{\sqrt{y^2+x^2+2xyt}}\sin\left( \frac{\pi
      y}{\sqrt{y^2+1}}\right)\cr 
    &\left.\quad\times\sin\left(\pi
      \sqrt{\frac{y^2+x^2+2xyt}{y^2+x^2+2xyt+1}}\right)\right]\cr  
    &\equiv\frac{4\pi\lambda}{N_c\bar{\rho}}
      \mathcal{I}\left(\frac{r}{\bar{\rho}}\right), 
\label{eq:int_pot}
\end{align}
where $y=z_I/\bar{\rho}$ and $t=\cos\theta$. Since it is inconvenient
and cumbersome to deal with Eq.~\eqref{eq:int_pot} in a practical
calculation, we introduce an interpolation function described in 
Refs.~\cite{Yakhshiev:2018juj,Musakhanov:2020hvk}. $\mathcal{I}$ in
Eq.~\eqref{eq:int_pot} is interpolated by the following form 
\begin{align}
    \mathcal{I}(x)=&I_0\left\{1+\sum_{i=1}^2\left[ a_i
   x^{2(i-1)}+a_3(-b_3 x)^i\right]e^{-b_i x^2}\right.\cr 
    &\left.+\frac{a_3}{x} \left(1-e^{-b_3 x^2}\right)\right\},
\end{align}
which is reduced to the asymptotic forms at $r=0$ and
$r\to\infty$ as follows 
\begin{align}
&\mathcal{I}(0)=0,\quad \mathcal{I}(\infty)=I_0
\end{align}
with the parameters $\mathcal{I}_0,\ a_i$ and $b_i$~\cite{Musakhanov:2020hvk}
\begin{align}
    &I_0=4.41625\cr
    &a=\left(\begin{array}{c}
    -1\\
    0.128702\\
    -1.1047
    \end{array}\right), 
    b=\left(\begin{array}{c}
        0.404875\\
        0.453923\\
        0.420733
        \end{array}\right).
\end{align}
As shown in Fig.~\ref{fig:2}, the interpolating function 
coincides with the numerical result for the original integral.
Using it, we can solve the Schr\"{o}dinger equation in an easy
manner. The Hamiltonian for the quarkonium system is written by 
\begin{align}
  H=-\frac{\hbar^2}{m_Q}\nabla^2+V_{Q\bar{Q}},
\end{align}
where $m_Q$ stands for the reduced mass of the heavy-quark and
$V_{Q\bar{Q}}$ denotes the $Q\bar{Q}$ potential that consists of 
\begin{align}
    V_{Q\bar{Q}}&=V_{\rm C}+V_I+V_{SD}+V_{SD}^I+V_{\rm s}.
    \label{eq:FullPot}
\end{align}
Here, $V_C$ and $V_s$ designate the Coulomb-like and confining
potentials, respectively. The linear potential (LP) has often been   
used for $V_s$~\cite{Godfrey:1985xj,Barnes:2005pb, 
Deng:2016stx,Yakhshiev:2016keg,Yakhshiev:2021jkc,PDSouza:2019kdx},
since it arises from the Wilson's area
law~\cite{Wilson:1974sk}. However, when a quark continues to separate 
from an antiquark, the string connecting them will be broken at a
certain scale and a light quark-antiquark pair will be created from
the vacuum. The LP does not explain this feature of the quark
confinement. This makes it difficult to explain the EM
transitions for excited states of the charmonia quantitatively. Thus,  
the strong screened potential for $V_s$ was used in
Refs.~\cite{Deng:2016stx}, which is saturated as $r$ increases. In the 
current work, we revise the strong screened potential by introducing
the Gaussian function, expressed as  
\begin{align}
    V_{\rm s} (r) &=  k(1-e^{-b r^2})/b,
\label{eq:confP}
\end{align}
which provides a better description of the EM transitions in the
presence of the instanton-induced heavy-quark potential. 
To solve the Schr\"{o}dinger equation, we use the variational
method that minimizes the eigenvalues with the six
experimental values of the charmonium masses used as input.
Minimizing the eigenvalues, we determine the fitting parameters listed in
Table~\ref{table:1}.  Model\,I is constructed by excluding the
instanton-induced potential, whereas Model\,II is built by including
it. 
\begin{table}[h]
    \caption{Fitting parameters to minimize the masses of the
      charmonia. We use the instanton parameters 
      ``$\bar{\rho}=1/3$\,fm and $R=1$\,fm."} 
  \begin{ruledtabular}
      \begin{tabular}{c|ccccc}
 Model &         $\alpha_s$(-) & $k$(GeV$^3$) & $\sigma$(GeV)
        &$m_c$(GeV)&$b$(GeV$^2$)\\ 
          \hline
Model\,I & 0.5016 & 0.0324 & 1.1895 & 1.5566 & 0.0244\\
Model\,II &          0.4863&0.0289&1.2112&1.5358&0.0234\\
      \end{tabular}
      \label{table:1}
  \end{ruledtabular}
  \end{table}
We want to emphasize that with the instanton-induced
potential and confining one given in Eq.~\eqref{eq:confP} we can use a 
value of the one-loop running strong coupling constant close to the
physical one at the charm-quark mass scale, which is written by
\begin{align}
  \alpha_s(\mu)=\frac{4\pi}{\beta_0}\frac{1}{\ln(\mu^2/
  \Lambda^2_{QCD})},
\end{align}
where $\beta_0=(11N_c-2N_f)/3$, $\Lambda_{\mathrm{QCD}}=
0.217$\,GeV\,\cite{Workman:2022ynf} and the scale $\mu\approx 
m_c$. If $\mu$ is fixed at the charm-quark mass $\mu\thickapprox
m_c=1.275$\,GeV, then $\alpha_s(\mu)=0.4258$.  
Note that this value is closer to that used in 
the instanton liquid model (see Table\,\ref{table:1}).

\begin{figure}[htbp]
  \centering
  \includegraphics[width=8cm]{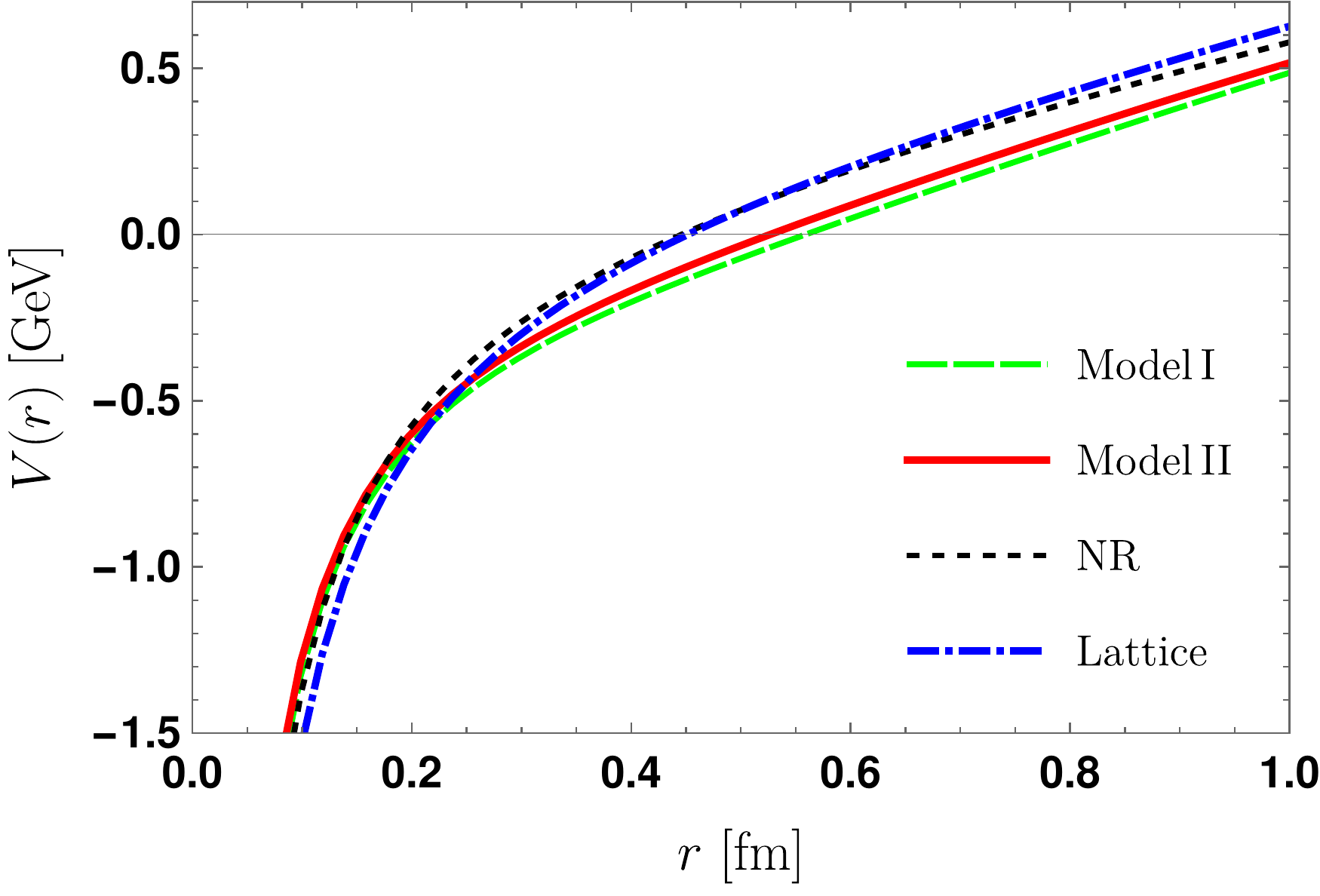}
  \caption{The central part of the $Q\bar{Q}$ potential.  
  The  green long-dashed and red solid curves draw the present
  results, whereas the black short-dashed and blue dash-dotted ones
  illustrate those from   the nonrelativistic model
  (NR)~\cite{Barnes:2005pb} and lattice QCD~\cite{Kawanai:2011jt},
  respectively.}  
  \label{figure:3}
\end{figure}

\begin{figure}[htbp]
  \centering
  \includegraphics[width=8cm]{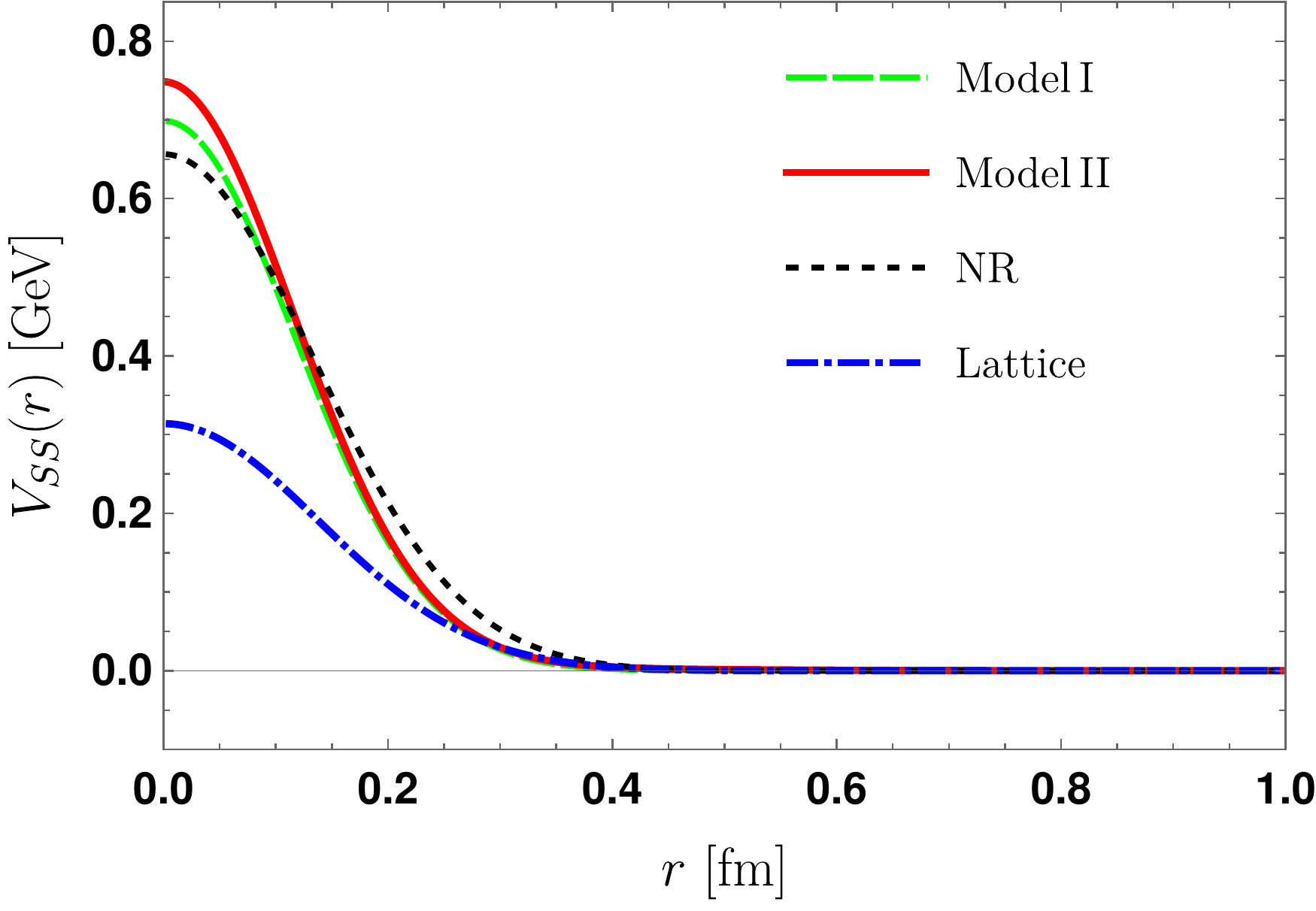}
  \caption{The spin-spin part of the $Q\bar{Q}$ potential. The
    notations are the same as in Fig.~\ref{figure:3}.} 
  \label{figure:4}
\end{figure}

It is of great interest to compare the heavy-quark potential with the
instanton effects to the phenomenological
potential~\cite{Barnes:2005pb} and that derived  
from lattice QCD~\cite{Kawanai:2011jt}. We draw the central and
spin-spin potentials respectively in Figs.~\ref{figure:3}
and~\ref{figure:4}. The potentials used for Model I and Model II are 
illustrated in the solid and dashed curves, whereas those from the NR
model and lattice QCD are depicted in the short-dashed and dot-dashed
ones, respectively.    
In Table~\ref{table:2}, the parameters used in the
NR approach~\cite{Barnes:2005pb} and  lattice
QCD~\cite{Kawanai:2011jt} are given in Table~\ref{table:2}.   
\begin{table}[htpb]
  \caption{The parameters of the $Q\bar{Q}$ potential in the NR
    model~\cite{Barnes:2005pb} and lattice QCD~\cite{Kawanai:2011jt}. } 
\begin{ruledtabular}
    \begin{tabular}{c|cccccc}
        &$\alpha_s$(-) & $\varkappa$(GeV$^2$) & $m_c$(GeV)&
    $\sigma$(GeV)&$A$(GeV)&$B$(GeV$^2$)\\  
        \hline
        NR&0.5461&0.1425&1.4794&1.0946&-&-\\
        LQCD&0.6010&0.1552&1.9888&1.0010&0.314&1.020
    \end{tabular}
    \label{table:2}
\end{ruledtabular}
\end{table}
Comparing the values of parameters listed in Table~\ref{table:2} with
those in Table~\ref{table:1}, we find that those of
$\alpha_s$ used in the NR approach and lattice QCD are larger than the
physical one employed in Model II. The charm-quark mass in the present
work is larger than that in Ref.~\cite{Barnes:2005pb} but is smaller
than in Ref.~\cite{Kawanai:2011jt}. 

The Coulomb-like potential from one-gluon exchange is
expressed as 
\begin{align}
  V_{\rm C}(r)=-\frac{4\alpha_s}{3r}.
\end{align}
In Model\,II, the contribution from the instanton vacuum is added (see
Eq.\,\eqref{eq:FullPot}). The confining potential employed in Model\,I
and Model\,II is defined in Eq.\,\eqref{eq:confP}, whereas the NR
model and lattice QCD use the linear type $V_{\rm s}=\varkappa r$.

In both Model\,I and the NR approach, the spin-spin potential given in 
Eq.\,\eqref{eq:VSS} are used, whereas in Model\,II we have
additionally Eq.\,\eqref{eq:VSSI}. On the other hand, in lattice QCD
the following form of the spin-spin potential is adopted
\begin{align}
  V_{SS}=A\exp(-B r^2).
  \label{eq:VSSLQCD}
\end{align}
We have extracted the values of $m_c$ and $\sigma$ in lattice QCD
by comparing Eqs.\,\eqref{eq:VSS} and\,\eqref{eq:VSSLQCD}. 

One can see that the central parts of all the four potentials in
Fig.~\ref{figure:3} are more or less the same at small distances. Only
at large distances, the potentials in Model I and II become slightly
lower than those used in the NR approach and Lattice QCD on account of 
the screening effect in Eq.\,\eqref{eq:confP}. The spin-spin part of
the present one is similar to that from the NR model but becomes very 
larger than that from lattice QCD.

As mentioned already, we use the six different values of the
charmonium masses as input, which are denoted by the asterisks in the
second column of Table~\ref{table:2}. Though, in general, the
instanton-induced potential provides a marginal contribution to the
masses of charmonia, they still improve the numerical results,
compared to the experimental data.  
\begin{table*}[h]
  \caption{Numerical results for the charmonium spectrum. The marked
    with the asterisk denotes the input data we have used to obtain
    the eigenvalues of the Schr\"{o}dinger equation. Model I and Model
    II represent the results of the present work without and with the
    instanton-induced potential. The columns with the NR and GI list
    respectively the results from the nonrelativistic linear potential
    and the Godfrey-Isgur relativized quark
    model~\cite{Barnes:2005pb}. Those with the LP and SP 
    are taken from Ref.~\cite{Deng:2016stx} in which both the linear
    confining potential and screened potential in the form
    of $(1-e^{-a r})$ are used, respectively.} 
  \begin{tabular}{c|c|c|c||c|c||c|c}
      \hline\hline
      State & Exp &  Model I & Model II &
    NR~\cite{Barnes:2005pb}&GI~\cite{Barnes:2005pb}&
    LP~\cite{Deng:2016stx}&SP~\cite{Deng:2016stx}\\ 
      \hline 
      $J/\psi (1^3S_1)$ \ &  $3096.900\pm0.006^*$
                  &3064&3097&3090&3098&3097&3097\\ 
      $\eta_c\ (1^1S_0)$ & $2983.9\pm0.4^*$&2957&2984&2982&2975&2984&2984\\
      $\psi\ (2^3S_1)$ & $3686.097\pm0.025^*$&3620&3687&3672&3676&3679&3679\\
      $\eta_c\ (2^1S_0)$ & $3637.6\pm1.1^*$&3572&3637&3630&3623&3635&3637\\
      $\psi\ (3^3S_1)$ & $4039\pm1$&4005&4089&4072&4100&4078&4030\\
      $\eta_c\ (3^1S_0)$ & &3976&4060&4043&4064&4048&4004\\
      $\psi\ (4^3S_1)$ & $4421\pm4$&4236&4338&4406&4450&4412&4281\\
     $\eta_c\ (4^1S_0)$ & &4223&4324&4384&4425&4388&4264\\
      $\chi_{c2}\ (1^3P_2)$ & $3556.17\pm0.07^*$&3493&3556&3556&3550&3552&3553\\
      $\chi_{c1}\ (1^3P_1)$ & $3510.67\pm0.05$& 3454&3511&3505&3510&3516&3521\\
      $\chi_{c0}\ (1^3P_0)$ & $3414.71\pm0.30$&3335&3414&3424&3445&3415&3415\\
      $h_c\ (1^1P_1)$ & $3525.38\pm0.11^*$&3467&3526&3516&3517&3522&3526\\
      $\chi_{c2}\ (2^3P_2)$ & $3922.5\pm1.0$&3913&3994&3972&3979&3967&3937\\
      $\chi_{c1}\ (2^3P_1)$ & $3871.65\pm0.06$&3879&3956&3925&3953&3937&3914\\
      $\chi_{c0}\ (2^3P_0)$ & $3862^{+26+40}_{-32-13}$&3784&3885&3852&3916&3869&3848\\
      $h_c\ (2^1P_1)$ & &3889&3967&3934&3956&3940&3916\\
      $\psi_3\ (1^3D_3)$ & $3842.7 \pm 0.2$&3746&3822&3806&3849&3811&3808\\
      $\psi_2\ (1^3D_2)$ & $3823.7\pm 0.5$&3743&3817&3800&3838&3807&3807\\
      $\psi\ (1^3D_1)$ & $3773.7\pm0.4$&3727&3799&3785&3819&3787&3792\\
      $\eta_{c2}\ (1^1D_2)$ & &3742&3817&3799&3837&3806&3805\\
      $\psi_3\ (2^3D_3)$ & &4092&4182&4167&4217&4172&4112\\
      $\psi_2\ (2^3D_2)$ & &4088&4177&4158&4208&4165&4109\\
      $\psi\ (2^3D_1)$ & $4191\pm5$ &4070&4157&4142&4194&4144&4095\\
      $\eta_{c2}\ (2^1D_2)$ & &4087&4177&4158&4208&4164&4108\\
      \hline\hline
  \end{tabular}
  \label{table:3}
\end{table*}

\section{The Electromagnetic Transitions of charmonia}
Since we have fixed the parameters for the heavy-quark potential by
using the mass spectrum of the charmonia, we are now in a position to
discuss the results for their EM transitions. 
The effective Hamiltonian for the quark-photon EM interaction is given
by 
\begin{align}
    H_{EM}=-e_c|e|\bar{\psi}\gamma^\mu A_\mu\psi,
\end{align}
where $\psi$ and $A_\mu$ denote the quark and photon field
operators. Using the EM Hamiltonian, we can compute the E1 and M1
transition matrix elements for the charmonia: 
\begin{align}
  \mathcal{M}_{\rm E1}^{fi} &=-i\omega_\gamma e_c|e|\langle
   f|\vec{r}\cdot\hat{\epsilon}^* e^{-i\vec{k}\cdot\vec{r}} |i\rangle,\cr
\mathcal{M}^{fi}_{\rm M1}& =-i\omega_\gamma
                              \frac{e_c|e|}{2m_Q}\langle
                              f|\vec{\sigma}\cdot(\hat{\epsilon}^*\times
                              \hat{k})e^{-i\vec{k}\cdot\vec{r}}|i\rangle ,
\end{align}
where the initial and final states are defined as 
\begin{align}
  |i\rangle&=|n;L,m_l;s,m_S;J,m_J\rangle\\
  |f\rangle&=|n';L',m_{l'};S',m_{s'};J',m_{J'}\rangle.
\end{align}

The E1 and M1 radiative partial widths are defined by
Ref.~\cite{Barnes:2005pb}: 
\begin{align}
    \Gamma_{\rm E1}&(n^{2S+1}L_J\rightarrow n'^{2S'+1}L'_{J'}+\gamma)\cr
    &=\frac{\omega_\gamma}{2\pi}|\langle f\,\gamma|\mathcal{M}_{\rm
      E1}|i\,0\rangle|^2\frac{E_f^{(c\bar{c})}}{M_i^{(c\bar{c})}}\cr 
    &=\frac{4}{3}C_{fi}\delta_{SS'}e_c^2\alpha|\langle
      \psi_f|r|\psi_i\rangle|^2
      E_\gamma^3\frac{E_f^{(c\bar{c})}}{M_i^{(c\bar{c})}},\label{eq:E1width}\\  
    \Gamma_{\rm M1}&(n^{2S+1}L_J\rightarrow
                     n'^{2S'+1}L_{J'}'+\gamma)\cr 
    &=\frac{\omega_\gamma}{2\pi}|\langle f\,\gamma|\mathcal{M}_{\rm
      M1}|i\,0\rangle|^2\frac{E_f^{(c\bar{c})}}{M_i^{(c\bar{c})}}\cr 
    &=\frac{4}{3}\frac{2J'+1}{2L+1}\delta_{LL'}\delta_{S,S'\pm 1}\cr 
    &\qquad\times
      e_c^2\frac{\alpha}{m_c^2}|\langle\psi_f|\psi_i\rangle|^2
E_\gamma^3\frac{E_f^{(c\bar{c})}}{M_i^{(c\bar{c})}},
\label{eq:M1width} 
\end{align}
where $E_\gamma$ represents the energy of the photon, 
$E_f^{(c\bar{c})}$ stands for the energy of the final $c\bar{c}$
state, $M_i^{(c\bar{c})}$ is the mass of the initial $c\bar{c}$ state,
and $C_{fi}$ is defined as   
\begin{align}
    C_{fi}={\rm max}(L,L')(2J'+1)\left\{
        \begin{array}{ccc}
            L' & J' & S\\
            J & L & 1
        \end{array}
    \right\}^2.
\end{align}
$\alpha$ denotes the fine-structure constant. $E_\gamma$ and
$E_f^{(c\bar{c})}$ are given as 
\begin{align}
    E_\gamma&=\frac{M_i^2-M_f^2}{2M_i},\cr
    E_f^{c\bar{c}}&=M_i-\frac{M_i^2-M_f^2}{2M_i}.
\end{align}
 In Eq.~\eqref{eq:E1width} and Eq.~\eqref{eq:M1width},
 $E_f^{c\bar{c}}/M_i^{c\bar{c}}$ is introduced as the relativistic
 phase space factor~\cite{Barnes:2005pb}. The wave functions for
the  charmonium states $\psi_i$ and $\psi_f$ are obtained by solving
the Schr\"odinger equation with the heavy-quark potential. 


\begin{table}[htp]
    \caption{Decay widths for 2S$\rightarrow$1P. 
Model I and Model II represent the results of the present work without
and with the instanton-induced potential. The columns with the NR and
GI list respectively the results from the nonrelativistic linear
potential and the Godfrey-Isgur relativized quark
model~\cite{Barnes:2005pb}. Those with the LP and SP 
    are taken from Ref.~\cite{Deng:2016stx} in which both the linear
    confining potential and screened potential in the form
    of $(1-e^{-a r})$ are used, respectively.
The decay rate is given in unit of keV. 
} 
\centering
\begin{tabular}{c|c|cc|cc|cc|c}
\hline\hline
\multicolumn{9}{c}{E1 transition}\\
\hline
\multirow{2}*{Initial} & \multirow{2}*{Final}&
\multicolumn{2}{c|}{Model}& 
\multicolumn{2}{c|}{\cite{Barnes:2005pb}}&
\multicolumn{2}{c|}{\cite{Deng:2016stx}}&PDG~\cite{Workman:2022ynf}\\ 
        \cline{3-9}
        &&I&II&NR&GI&LP&SP&Exp.\\ 
        \hline
        \multirow{3}*{$\psi(2^3S_1)$}& $\chi_{c2}(1^3P_2)$
       &40&41&38&24&36&44&$28 \pm 1$\\
         & $\chi_{c1}(1^3P_1)$&42 &43&54&29&45&48&$29 \pm 1$\\
        & $\chi_{c0}(1^3P_0)$&28&28&63&26&27&26&$29 \pm 1$\\
        $\eta_c(2^1S_0)$ & $h_c(1^1P_1)$&43 &41&49 &36 &49&52&-\\
        \hline\hline
    \end{tabular}
    \label{table:4}
\end{table}
In Table~\ref{table:4}, we list the results for the E1 transition from
the 2S to 1P states. As shown from those listed in the third and
fourth columns, the instanton effects reduce the strengths of the
decay rates for $\psi(2^3 S_1)\to \chi_{c1,c2}(1^3 P_{0,1,2})$ but the
results are still larger than the experimental
data~\cite{Workman:2022ynf}. On the other hand, they do not contribute
to its decay to $\chi_{c0}$ and $h_c$ almost at all and are in good
agreement with the data.   
 
\begin{table}[htp]
\caption{Decay widths for 1P$\rightarrow$1S. 
Notations are the same as in Table~\ref{table:4}.
} 
\centering
\begin{tabular}{c|c|cc|cc|cc|c}
\hline\hline
\multicolumn{9}{c}{E1 transition}\\
\hline
\multirow{2}*{Initial} & \multirow{2}*{Final}& 
\multicolumn{2}{c|}{Model}& \multicolumn{2}{c|}{\cite{Barnes:2005pb}}
&\multicolumn{2}{c|}{\cite{Deng:2016stx}}&PDG~\cite{Workman:2022ynf}\\
\cline{3-9}
&&I&II&NR&GI&LP&SP&Exp.\\ 
\hline
$\chi_{c2}(1P)$&\multirow{3}{*}{$J/\psi(1S)$}&388&394
&424&313&327&338&$374\pm19$ \\
$\chi_{c1}(1P)$&&311&315&314&239&269&278&$288 \pm 16$\\
$\chi_{c0}(1P)$&&146&152&152&114&141&146&$151 \pm 12$\\
$h_c(1P)$& $\eta_c(1S)$ &445&452& 498 & 352 & 361&373&$350 \pm 210$ \\
\hline\hline
\end{tabular}
\label{table:5}
\end{table}
Table~\ref{table:5} lists the results for the E1 transition from the
1P to 1S states. In contrast to the E1 transitions for $\mathrm{2S}\to
\mathrm{1P}$ decays, the instanton effects slightly enhance the decay 
widths. The decay widths for the $\chi_{cJ}(1P)\to J/\psi(1S)$ are well
described by the present heavy-quark potential. That for the
$h_c(1P)\to \eta_c(1S)$ is overestimated, compared with the data. 

\begin{table}[htp]
    \caption{Decay widths for 1D$\rightarrow$1P. 
Notations are the same as in Table~\ref{table:4}.
} 
        \centering
        \begin{tabular}{c|c|cc|cc|cc|c}
            \hline\hline
            \multicolumn{9}{c}{E1 transition}\\
            \hline
            \multirow{2}*{Initial} & \multirow{2}*{Final}& 
    \multicolumn{2}{c|}{Model}& 
    \multicolumn{2}{c|}{\cite{Barnes:2005pb}}&
    \multicolumn{2}{c|}{\cite{Deng:2016stx}}&
    PDG~\cite{Workman:2022ynf}\\
            \cline{3-9}
            &&I&II&NR&GI&LP&SP&Exp.\\ 
            \hline
            $\psi_3(1^3D_3)$ & \multirow{2}*{$\chi_{c2}(1^3P_2)$} 
    &331&333&272&296&377&393 &-\\
            $\psi_2(1^3D_2)$ & &80 &80&64&66&79&82 &-\\
             & $\chi_{c1}(1^3P_1)$&311 &311&307&268&281&291&- \\ 
            \multirow{3}*{$\psi(1^3D_1)$}& $\chi_{c2}(1^3P_2)$ 
    &7.2&7.2&4.9&3.3&5.4&5.7&$\leq 17$\\
            & $\chi_{c1}(1^3P_1)$&147&147&125&77&115&111&$68 \pm 7$\\
            & $\chi_{c0}(1^3P_0)$&317 &315&403&213&243&232&$188 \pm 18$ \\
            \hline\hline
        \end{tabular}
        \label{table:6}
\end{table}
In Table~\ref{table:6}, we present the results for the $1\rm
D\rightarrow 1 P$ decay widths. The decay rates for the $\psi(1^3
D_1)\to \chi_{c0,c1}(1^3P_{0,1})$ are overestimated, compared with
the data. The instanton effects are again very small. 
Note that the $\psi(3770)$ resonance lies just above the 
open-charm $D\bar{D}$ threshold. It implies that the light quarks may
come into play. Furthermore, the $1^3D_1$ state $\psi(3770)$
is assumed to be mixed with a small $2^3 S_1$
state~\cite{Rosner:2001nm, Rosner:2004wy}. One can improve the
results by considering the light-quark contributions and $S-D$
mixing. Since the $\psi(3770)$ is above the $D\bar{D}$ threshold,
relativistic corrections may also be important. However, since we aim at
the effects of the heavy-quark potential from the instanton vacuum in
the current work, we have not considered taken them.

\begin{table}[htp]
    \caption{Decay widths for the M1 transition. 
Notations are the same as in Table~\ref{table:4}.
}
        \centering
        \begin{tabular}{c|c|cc|cc|cc|c}
            \hline\hline
            \multicolumn{9}{c}{M1 transition}\\
            \hline
            \multirow{2}*{Initial} & \multirow{2}*{Final}& 
    \multicolumn{2}{c|}{Model}& \multicolumn{2}{c|}{\cite{Barnes:2005pb}}&
    \multicolumn{2}{c|}{\cite{Deng:2016stx}}&PDG~\cite{Workman:2022ynf}\\
            \cline{3-9}
            &&I&II&NR&GI&LP&SP&Exp.\\ 
            \hline
            $J/\psi(1S)$&$\eta_c(1S)$&2.5&2.3&2.9&2.4&2.39&2.44
    &$1.7 \pm 0.4$\\
            $\psi(2S)$&$\eta_c(2S)$&0.2&0.2&0.21&0.17&0.19&0.19
    &$0.2 \pm 0.2$\\
             &$\eta_c(1S)$&5.4&5.0&4.6&9.6&8.08&7.80&$1.0 \pm 0.1$\\
            $\eta_c(2S)$&$J/\psi(1S)$&9.6&9.0&7.9&5.6&2.64&2.29&$< 158$\\
            \hline\hline
        \end{tabular}
        \label{table:7}
\end{table}
In Table~\ref{table:7}, we list the results for the decay rates for
the M1 transition. Though the results describe the
experimental data well, the instanton effects are again negligibly
small. 

\section{Summary}
In the current work, we examined the instanton effects on the mass
spectrum and electromagnetic transitions of charmonia. We briefly 
reviewed how the heavy-quark potential from the instanton vacuum. To
improve the results, we introduced a new type of the confining
potential. We utilized the experimental data for the masses of the
six different charmonia to fit the data. 
Though the instanton effects are marginal on the mass
spectrum of the charmonia, they allow one to use smaller values of the
strong coupling constant and the charm-quark mass, which are closer to
the physical values compared to other works. Using the effective
Hamiltonian for the quark-photon electromagnetic interaction, we
derived the radiative decay rates for the E1 and M1 transitions. The
instanton heavy-quark potential has generally minor effects on the
radiative decays of the charmonia. 

\section*{Acknowledgments}
The present work was supported by Basic Science Research Program
through the National Research Foundation of Korea funded by the
Ministry of Education, Science and Technology
(Grant-No. 2021R1A2C209336 and 2018R1A5A1025563 (H.-Ch. K.), and
2020R1F1A1067876 (U. Y.)).

\end{document}